# *Privacy Preservating Gen AI*

Mr.U Thiruvaazhi
Information Science and Engineering,
Kumaraguru College of Technology,
Coimbatore, Tamil Nadu
thiruvaazhi.u.ise@kct.ac.in

Balavedhaa S
Artificial Intelligence and Data Science
Kumaraguru College Of Technology
Coimbatore, Tamil Nadu
balavedhaa.21ad@kct.ac.in

Ganesh R
Computer Science and Engineering
Kumaraguru College Of Technology
Coimbatore, Tamil Nadu
ganesh.21cs@kct.ac.in

Rakshana M
Information Science and Engineering,
Kumaraguru College of Technology,
Coimbatore, Tamil Nadu
rakshana.21is@kct.ac.in

Ram Sundhar K Shaju
Information Science and Engineering,
Kumaraguru College of Technology,
Coimbatore, Tamil Nadu
ramsundhar.21is@kct.ac.in

Swetha S
Information Science and Engineering,
Kumaraguru College of Technology,
Coimbatore, Tamil Nadu
swetha.21is@kct.ac.in

*Abstract*—The ability of machines to comprehend and produce language that is similar to that of humans has revolutionized sectors like customer service, healthcare, and finance thanks to the quick advances in Natural Language Processing (NLP), which are fueled by Generative Artificial Intelligence (AI) and Large Language Models (LLMs). However, because LLMs trained on large datasets may unintentionally absorb and reveal Personally Identifiable Information (PII) from user interactions, these capabilities also raise serious privacy concerns. Deep neural networks' intricacy makes it difficult to track down or stop the inadvertent storing and release of private information, which raises serious concerns about the privacy and security of AI-driven data. This study tackles these issues by detecting Generative AI weaknesses through attacks such as data extraction, model inversion, and membership inference. A privacy-preserving Generative AI application that is resistant to these assaults is then developed. It ensures privacy without sacrificing functionality by using methods to identify, alter, or remove PII before to dealing with LLMs. In order to determine how well cloud platforms like Microsoft Azure, Google Cloud, and AWS provide privacy tools for protecting AI applications, the study also examines these technologies. In the end, this study offers a fundamental privacy paradigm for generative AI systems, focusing on data security and moral AI implementation, and opening the door to a more secure and conscientious use of these tools.

## I. Introduction

Large Language Models (LLMs) like GPT and BERT, in particular, have revolutionized industries including education, health care, finance, and customer service by producing language that is human-like by training on enormous datasets. These models facilitate a variety of uses, such as real-time communication, natural language comprehension, and automated content creation. But there are serious privacy issues with their increasing use, particularly with regard to the inadvertent exposure and keeping of Personally Identifiable Information (PII) that is incorporated into training data. The inability to track or regulate the extent to which sensitive data may be revealed due to these models' black-box nature causes users to hesitate and raises the need for privacy-preserving solutions.

By examining the privacy risks of Generative AI, including membership inference, model inversion, and data extraction, and calculating their impact and probability using structures like the Data Sensitivity-Exposure (DSE) Model, in accordance with NIST SP 800-30 and ISO/IEC 27005, this project tackles these issues. Depending on the model's vulnerability and the sensitivity of the data, risks are classified as low, medium, or high. Individuals may experience identity theft or fraudulent activity as a result of privacy breaches, while businesses may experience financial loss, harm to their brand, and legal repercussions. To guarantee ethical and secure use, strong countermeasures such as data anonymization, differential privacy, and more stringent data management procedures are necessary.

Generative AI systems must comply with laws like the GDPR, HIPAA, CCPA, and India's PDP Bill in order to maintain trust and compliance. These regulations require transparency, user control, and stringent guidelines for handling personal data. In order to provide a comprehensive framework that improves data protection without lowering the usefulness of AI models, the project investigates and contrasts the privacy solutions now provided by cloud providers such as Microsoft Azure, Google Cloud, and AWS. In order to facilitate the safe implementation of generative AI technologies, next chapters will go into detail on studies of literature, experimental approaches, and the assessment of different privacy-enhancing strategies

## II. Literature

As LLM06 in the OWASP LLM Top 10, the possibility of sensitive data leaking through LLMs has grown to be a serious worry. Because these models were trained on large datasets, they may inadvertently reveal sensitive information like healthcare records, proprietary company material or personally identifiable information (PII). Direct output reproduction, deftly constructed user prompts or sophisticated attacks like model inversion and membership inference can all result in such revelations. The data used for training itself, which may include confidential data that the model subsequently recalls, is frequently the source of the problem.

Recent studies and OWASP suggest a number of technical precautions to lessen these dangers. Typical strategies include using adversarial training techniques, using differential privacy, putting access restriction mechanisms in place and screening pre-training data. More sophisticated techniques, such as homomorphic encryption, federated learning, secure multi-party computation and data anonymization, seek to preserve model accuracy and usefulness while safeguarding privacy. These are particularly important in high-risk industries where security of information is required by law and ethics, such healthcare and banking.

Alongside model architectures, privacy-preserving methods are also developing. A promising technique that lessens the possibility of memorization-based leaks is Retrieval-Augmented Generation (RAG), which enables models to consult external knowledge sources rather than depending only on internal training data. Advanced models like Zephyr 7B, which are based on the Mistral-7B framework and optimized using Direct Preference Optimization (DPO), show advancements in conversational AI, but they also emphasize how crucial it is to incorporate privacy safeguards into new developments. In the end, creating safe, open, and reliable AI systems requires coordinating technological advancements with moral principles.

### III. PRIVACY VULNERABILITY INSTANCES

In the quickly developing fields of AI and machine learning, protecting data privacy has grown in complexity and importance. Models are increasingly susceptible to a range of privacy threats that can jeopardize the security of both individuals and organizations as they become more complex and undergo training on enormous datasets, many of which contain private or sensitive data. Membership inference attacks are one such risk, in which a malicious party can ascertain if a specific data point was utilized during training, thus disclosing personal user information. Even more concerning are model inversion attacks, which use the model's predictions to reconstitute the actual input data—like medical records or face photos. These risks demonstrate how seemingly harmless model results may inadvertently disclose private data.

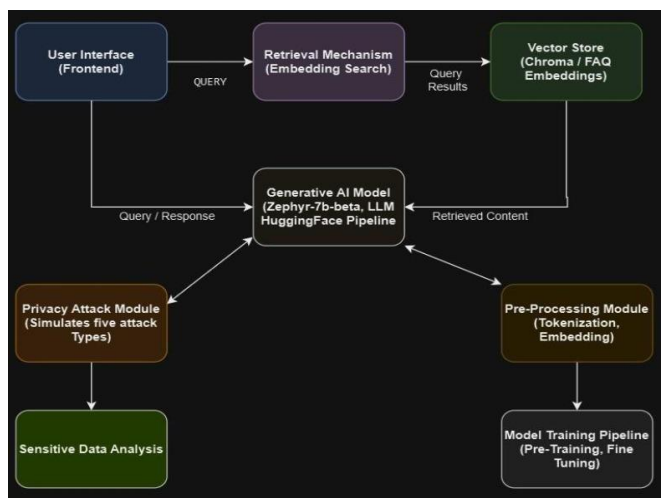

Data extraction attacks, which allow adversaries to retrieve specific details stored in the model—sometimes exposing proprietary data or personally identifiable information (PII)—increase the danger even further. By removing high-level traits or statistical associations from training data, property inference attacks adopt a more comprehensive strategy and expose private group-level trends such as political affiliations, economic brackets or health issues.

#### A. Membership Inference Attack

The attacks using membership inference examine a model's response to particular inputs to ascertain whether or not those inputs were included in the training set. The discovery that models such as Variational Autoencoders (VAEs) frequently recreate known (training) information more accurately than unseen data serves as the foundation for this. We used a fake financial dataset created with the Faker package in our setup, which included private information such as email addresses, SSNs, and credit card numbers. This data was encoded and reconstructed using a VAE. Reconstruction errors are then calculated by the attacker; if the error falls below a certain level, the data is assumed to be a part of the data set used for training. This technique demonstrates how model behavior may inadvertently reveal membership details.

Particularly in industries where confidentiality of information is crucial, like finance and healthcare, such attacks have serious ramifications. If successful, attackers can find and abuse private data linked to specific people, resulting in privacy violations. Overfitting, in which the model retains training data rather than generalizing, is the main cause of this weakness. Attackers can determine the information that the model was trained on by identifying these patterns. Given that even indirectly model outputs have the potential to jeopardize personal information, this emphasizes the necessity of privacy-preserving strategies in AI development.

#### B. Data Extraction Attack

Data extraction attacks are attempts to retrieve sensitive information that a trained model may have learned from its data used for training by providing it with carefully constructed inputs. This frequently happens when the model, particularly big language models like GPT-2, unintentionally saves and repeats personal data like names, contact information or email addresses. We employed a synthetic dataset in our experiment that contained information about made-up people like Alice and John. After the model was adjusted using this data, prompts aimed at particular attributes were used to query it. The model suggested possible recollection and privacy leakage if it provided corresponding personal information.

These assaults carry significant ramifications, especially in regulated industries like e-commerce, banking and healthcare. They demonstrate how models may inadvertently make training data public, which could result in identity theft or other privacy violations. The primary cause of this privacy loss is model memorization, which preserves delicate patterns rather than generalizing them. The risk is increased when privacy-preserving strategies, such as differential privacy are not used. Protecting training data from extraction is crucial to preserving confidence and safeguarding user information as AI models continues to expand in scope and use.

### C. Data Poisoning Attack

Data poisoning attacks are deliberate attempts to alter a model's behavior by inserting tainted or false data into the training set. An attacker could, for example, mark emails that resemble spam as "non-spam" in order to trick a classifier and reduce its accuracy and dependability. In order to mimic sensitive information like names and addresses, we created a fake financial dataset in our experiment using the Faker package. Based on characteristics like age and wealth, a Feedforward Neural Network (FNN) was trained to forecast binary outcomes. With the intention of biasing the model and causing privacy violations or misclassifications, the attacker covertly tainted 50% of data samples by introducing noise and inverting labels.

These assaults jeopardize privacy in addition to performance. Under specific searches, models may memorize and repeat personal information such as location or salary due to poisoned data. This could potentially lead to inadvertent PII exposure or produce biased results in delicate fields like healthcare or finance. A decrease in accuracy and evidence of information leakage were found when clean and poisoned models were compared. Poisoning models can compromise user trust, disclose sensitive data and create opportunities for additional attacks such model inversion if adequate data validation or differential privacy aren't in place.

### D. Model Inversion Attack

When an attacker tries to deduce sensitive input characteristics, such as age or gender, only from a model's outputs, this is known as a model inversion attack. In industries such as healthcare, where models handle Protected Health Information (PHI), this type of privacy violation is very worrisome. In order to predict the chance of sickness, we trained a fully connected neural network (FCNN) in PyTorch using a synthetic medical dataset that had the following fundamental attributes: age, gender and condition. Then, utilising the Binary Cross-Entropy loss and Adam optimizer, the attacker executes an inversion attack by beginning with random age and gender predictions and iteratively updating them to match a particular model result.

The optimizer effectively reconstructs inputs that could have produced the observed result by progressively adjusting the estimated inputs over 1000 steps in order to minimize the loss. This black-box querying procedure demonstrates how, even in the absence of immediate access to the training dataset, model outputs can reveal private information. There are serious repercussions because attackers might retrieve personal information such as gender and health issues, endangering patient privacy and violating confidentially. These assaults highlight the critical need for strong defenses, moral AI methods and secure deployment mechanisms to safeguard sensitive data. They are caused by model overexposure and a lack of privacy-preserving strategies like differential privacy.

### E. Property Inference Attack

In a property inference attack, the attacker strives to deduce statistical variables from the data set used for training a model without obtaining specific PII. A model, like GPT-3.5 or GPT-4, that has been optimized using data containing sensitive attributes, such as email addresses or demographic characteristics, is trained or queried by the attacker. The objective is to infer patterns, such as the probability of a location, name, or identification showing up in the training data, rather than to find specific information entries. For instance, based on its responses to prompts, the model can still imply the existence of email addresses even if it is not aware of any real ones. Because it exposes latent dataset properties and violates standards of privacy and legislation, this leakage is particularly risky when models are developed on datasets that contain personally identifiable information.

In the experimental setting, we create an artificial financial dataset with names, emails, SSNs, credit card information and more using the Faker package. This data is then used to fine-tune a previously trained language model that encodes the dataset's statistical characteristics. To check for biases, the attacker creates prompt patterns, such as prompting the model to fill in incomplete information or asking about demographics. Even without explicit access to the training data, the attacker can identify indications of latent property learning by comparing outcomes across various prompts. In industries where disclosing statistical distributions (such as categories of wealth or medical issues) can have major repercussions, such as banking, healthcare and law, these attacks are especially damaging.

## IV. PRIVACY AWARE AI FRAMEWORK

The technical stack that supports the AI-powered data processing methodology is safe, scalable, and privacy-conscious. To prevent unwanted access and lessen Denial-of-Service (DoS) attacks, API security is implemented by rate limitation, role-based access control (RBAC) and key-based authentication. AES-256 is used for data encryption at rest, and HTTPS is used for data encryption in transit. Encryption keys are protected by an encrypted key management system. Both organized and unorganized sources of data are collected and they are then run through a preliminary processing pipeline that includes standardization, tokenization, and normalization. In order to facilitate effective similarity-based retrieval, the processed data is further transformed into embeddings using sophisticated language models and saved in the Weaviate vector database.

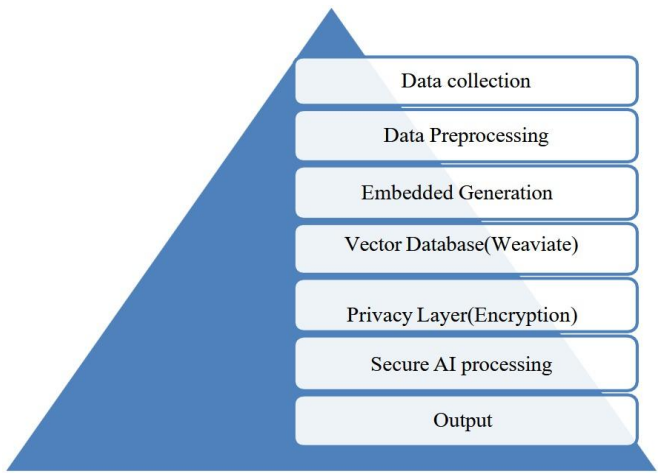

The architecture incorporates a robust privacy layer using methods including the use of tokens masking, generalizations and differential privacy to safeguard sensitive data. These techniques guarantee the concealment of personally identifiable information (PII) while maintaining analytical utility. Deployment makes use of cloud platforms like Scaleway for flexible, affordable deployment and Azure OpenAI for safe model hosting and API management. Scalability and automation are supported via Infrastructure as Code (IaC). A comparison of the two cloud platforms aids in performance, price and security optimization. The layered architecture as a whole guarantees reliable, private and efficient AI-based data processing.

A wide range of criteria, with an emphasis on effectiveness, efficiency and security, are used to assess the efficacy of the privacy-preserving AI system. Embedding efficiency gauges the caliber of representations of vectors that provide these results, whereas retrieval accuracy shows how well the system delivers pertinent information. Scalability guarantees that the framework operates reliably under increasing data and user demands, while Latency offers insight into system responsiveness. Privacy risk measures the degree to which the framework employs privacy-preserving strategies to safeguard sensitive data. When combined, these criteria provide a comprehensive evaluation of data confidentiality and technological performance.

The architecture incorporates a robust privacy layer using methods including the use of tokens masking, generalizations and differential privacy to safeguard sensitive data. These techniques guarantee the concealment of personally identifiable information (PII) while maintaining analytical utility. Deployment makes use of cloud platforms like Scaleway for flexible, affordable deployment and Azure OpenAI for safe model hosting and API management. Scalability and automation are supported via Infrastructure as Code (IaC). A comparison of the two cloud platforms aids in performance, price and security optimization. The layered architecture as a whole guarantees reliable, private and efficient AI-based data processing.

A wide range of criteria, with an emphasis on effectiveness, efficiency and security, are used to assess the efficacy of the privacy-preserving AI system. Embedding efficiency gauges the caliber of representations of vectors that provide these results, whereas retrieval accuracy shows how well the system delivers pertinent information. Scalability guarantees that the framework operates reliably under increasing data and user demands, while Latency offers insight into system responsiveness. Privacy risk measures the degree to which the framework employs privacy-preserving strategies to safeguard sensitive data. When combined, these criteria provide a comprehensive evaluation of data confidentiality and technological performance.

## V. RESULT AND CONCLUSION

This chapter examines five major privacy flaws in generative AI systems that pose significant dangers to security and privacy: membership inference, model inversion, property inference, data extraction and data poisoning attacks. It highlights the processes underlying these incidents and the urgent need for strong privacy-preserving techniques in AI research through experimental examples. The incorporation of these methods into a framework for AI-powered data processing is also covered, emphasizing vector databases, cloud-based deployment and safe model building as essential components. These methods serve as the foundation for systems that are both scalable and privacy-conscious. The chapter lays the groundwork for upcoming talks about performance reviews and practical applications.